%% file: FPCP2012-40.tex
\documentclass[floatfix,twocolumn,twoside,slac_two]{revtex4}
\usepackage{graphicx}
\usepackage{fancyhdr}
\usepackage{microtype}
\usepackage{hepparticles}
\usepackage{hyperref}
\usepackage{hepnicenames}
\usepackage{hepunits}
\usepackage{feynmp}
\usepackage{subfig}
\begin{document}

\title{Searches for Lepton Flavour Violation and Lepton Number Violation in Hadron Decays}

\author{P. Seyfert on behalf of the LHCb Collaboration}
\affiliation{Physikalisches Institut, Heidelberg University, Im Neuenheimer Feld 226, 69120 Heidelberg, Germany}

\begin{abstract}

In the Standard Model of particle physics, lepton flavour and lepton number are
conserved quantities although no fundamental symmetry demands their
conservation. I present recent results of searches for lepton flavour and
lepton number violating hadron decays measured at the $\PB$ factories and LHCb.

In addition, the LHCb collaboration has recently performed a search for the
lepton flavour violating decay $\Ptauon\to\Pmuon\Pmuon\APmuon$. The obtained
upper exclusion limit, that has been presented in this talk for the first time,
is of the same order of magnitude as those observed at the $\PB$ factories.
This is the first search for a lepton flavour violating $\Ptau$ decay at a
hadron collider. 
\end{abstract}

\maketitle

\thispagestyle{fancy}

\input{content}

\input{FPCP2012-40.bbl}
\end{document}

%% file: content.tex
\section{Introduction}

In the Standard Model both lepton number as well as lepton flavour are
conserved quantities \cite{SM1,SM2}. Since both can be broken in extensions of
the standard model, observation of either of them would be a clear sign for new
physics.

Results of the search for lepton number violation (LNV) and lepton flavour
violation (LFV) in decays of hadrons are presented. These comprise the $\PB$
decay modes $\PBplus\to h^+\ell^+\ell^{\prime -}$, $\PBplus\to
h^-\ell^+\ell^{\prime +}$, and the corresponding $\PD$ modes $\PDplus\to
h^+\ell^+\ell^{\prime -}$ and $\PDplus\to h^-\ell^+\ell^{\prime +}$.  The final
state meson $h$ may hereby either be a stable meson (pions or kaons) or in case
of $\PB$ decays also a $\PD$ meson. Additionally to these modes, the first
limit on the branching fraction $\PBplus\to\PDzero\Ppiplus\Pmuon\Pmuon$ is
presented as well as a new result on the search for LFV in the decay of $\Ptau$
leptons at the LHC.

Throughout this document charge conjugate decays are implied.

\subsection{Lepton Number Violation}

\begin{figure}
\centering
\subfloat[][on-shell Majorana neutrino]{\includegraphics[width=.45\textwidth]{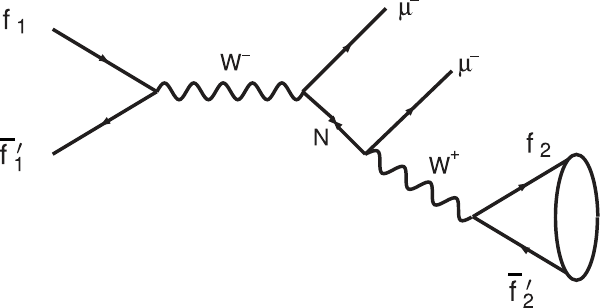}}
\hspace{.05\textwidth}
\subfloat[][virtual Majorana neutrino]{\includegraphics[width=.28\textwidth]{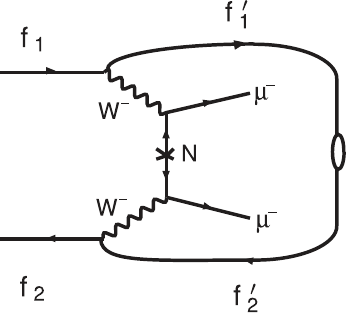}}
\caption{Lowest order diagrams for LNV in meson decays, involving on-shell or virtual Majorana neutrinos. Depending on the individual quark flavours either of them can be Cabbibo favoured. Reproduced from \cite{lhcb-Blnv2}.}
\label{fig:onshellvsoffshell}
\end{figure}

Numerous models without lepton number conservation have been proposed, see
\cite{majorana} for an overview.  Similar to the fundamental diagram in the
neutrinoless double beta decay, any neutrinoless hadron decay with two same
sign leptons in the final state probes the existence of Majorana neutrinos.  Of
the two lowest order diagrams for LNV in meson decays, one can go through an
on-shell neutrino, while the other contains a virtual neutrino
(Fig.~\ref{fig:onshellvsoffshell}). In $\PBplus$ decays one of them is Cabbibo
favoured depending on the final state. Thus for Majorana neutrino masses in the
accessible mass range (up to $\unit{5140}{\MeVovercsq}$) the modes
$\PBplus\to\PDminus\Pleptonplus\Pleptonplus$ and
$\PBplus\to\PDstar^{-}\Pleptonplus\Pleptonplus$ are more sensitive and also
provide a mass measurement. Beyond the accessible mass range other final states
($\Ppiminus\Pleptonplus\Pleptonplus$, $\PDsminus\Pleptonplus\Pleptonplus$) are
more sensitive.

\begin{figure}
\centering
\subfloat[][coupling to electrons]{\includegraphics[width=.45\textwidth]{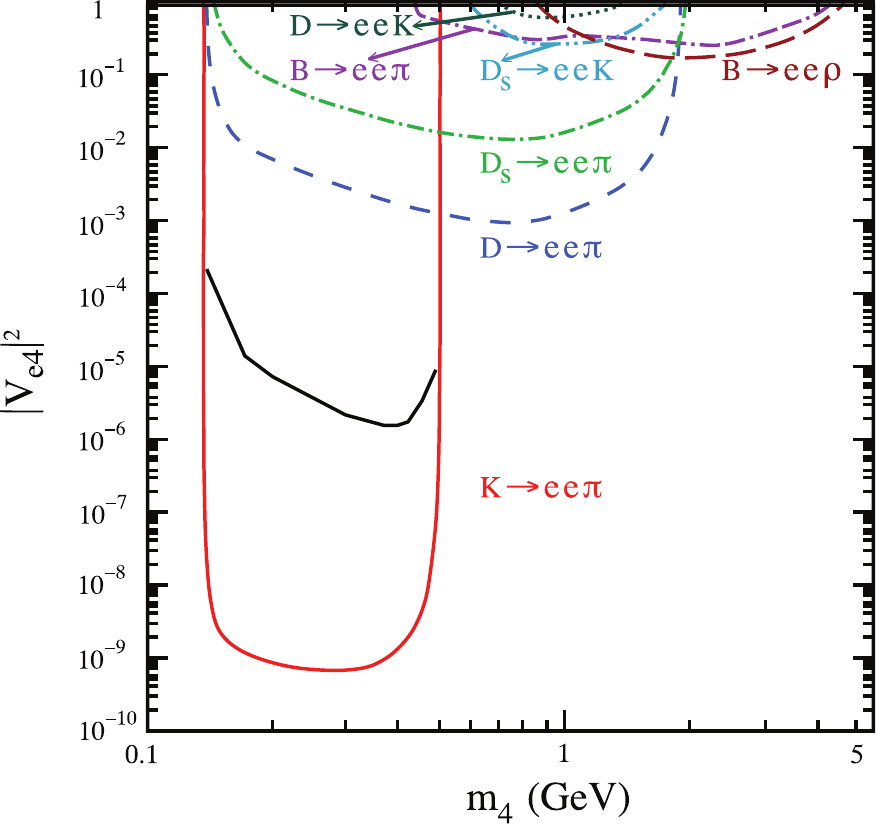}}
\hspace{.05\textwidth}
\subfloat[][coupling to muons]{\includegraphics[width=.45\textwidth]{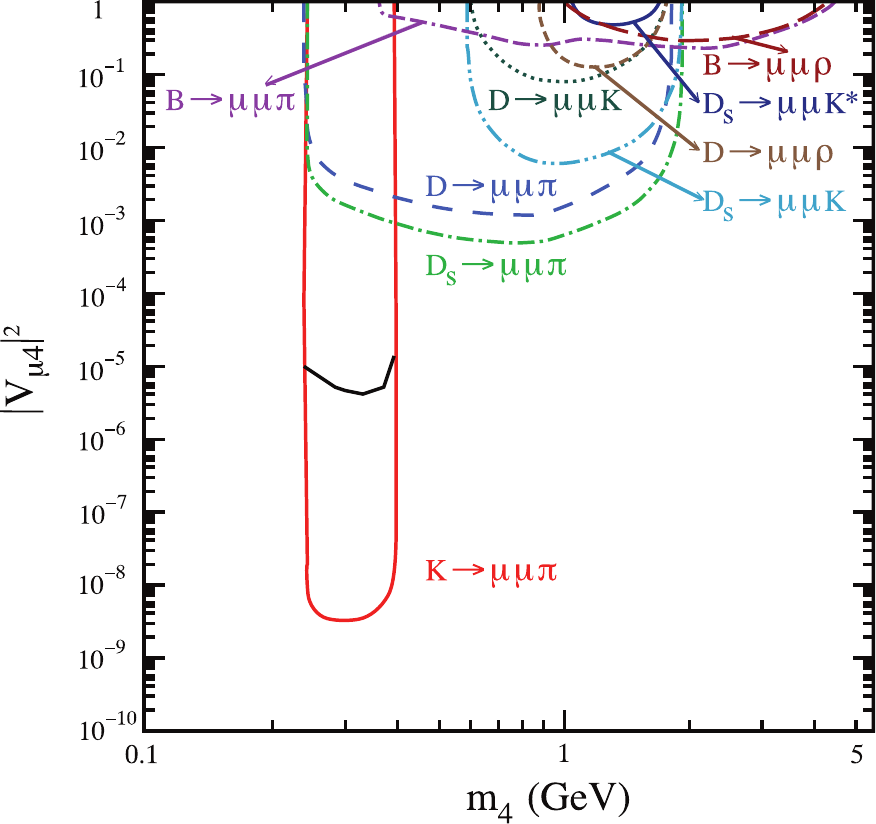}}
\caption{Constraints on charged lepton couplings $V_{\Pe 4}$ and $V_{\Pmu 4}$ to a fourth heavy Majorana neutrino from 2009 as a function of the mass $m_4$\cite{majorana}.}
\label{fig:majoranaoverview}
\end{figure}

In the framework of LNV through a fourth neutrino $N$ with a large Majorana
mass, an observation of LNV not only provides information about the mass $m_4$
of the fourth neutrino, but also on the $\PW N\ell$ coupling strength
$|V_{\Plepton 4}|$. A compilation of different exclusion limits is shown in
Fig.~\ref{fig:majoranaoverview}. For the coupling to the muon the strongest
constraints come from kaon physics.

\begin{figure}
\centering
\includegraphics[width=.45\textwidth]{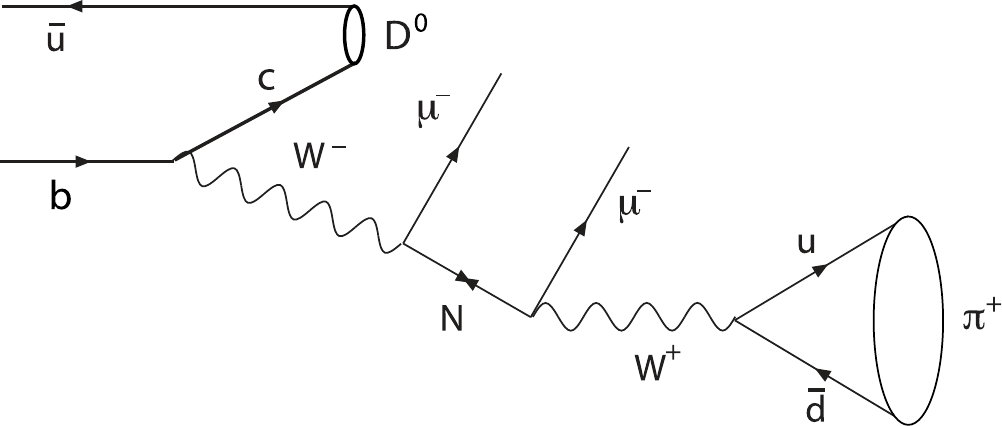}
\caption{Tree level Feynman diagram for the decay $\PBminus\to\PDzero\Ppiplus\Pmuon\Pmuon$.}
\label{fig:threebody}
\end{figure}

Complementary to the modes with one meson in the final state, it has been
suggested in \cite{1108.6009} to also consider
$\PBplus\to\PDzero\Ppiplus\Pmuon\Pmuon$ with the diagram shown in
Fig.~\ref{fig:threebody}. Until 2012 no limit on the branching fraction of this
decay has been measured.

\subsection{Lepton Flavour Violation}

\begin{figure}
\centering
\includegraphics[width=.45\textwidth]{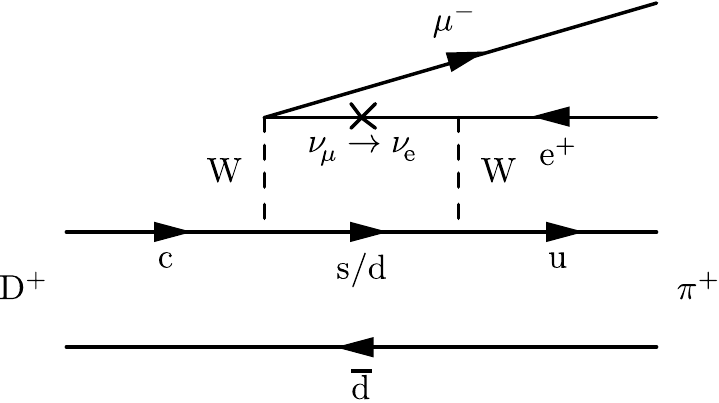}
\caption{Feynman diagram for lepton flavour violating meson decays in the Standard Model with neutrino oscillation.}
\label{fig:hadlfvsm}
\end{figure}

In contrast to violation of lepton number, lepton flavour violation has been
observed experimentally in the neutrino sector. Through loop diagrams, neutrino
oscillation can also enter the charged sector as illustrated in
Fig.~\ref{fig:hadlfvsm} -- the predicted rates however are immeasurable small,
suppressed by powers of $m^2_{\Pneutrino}/m^2_{\PW}$ \cite{lfvcalc}.

\begin{figure}
\centering
\subfloat[][Standard Model extension with a new scalar boson $\Phi$ \label{fig:phis}]{\includegraphics[width=.45\textwidth]{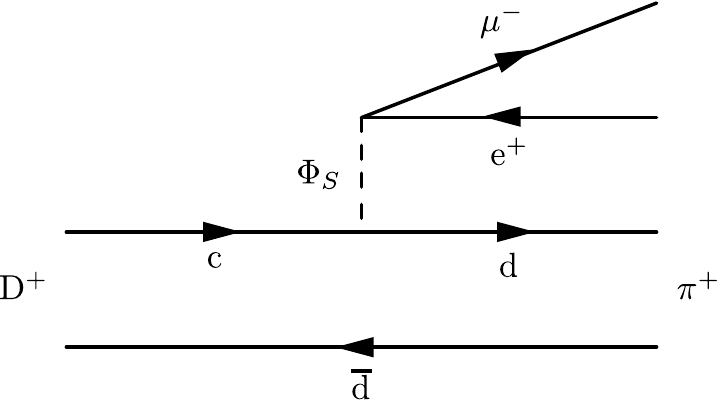}}
\hspace{.05\textwidth}
\subfloat[][Standard Model extension with a heavy neutrino $\nu_H$\label{fig:hadlfvbsmtwo}]{\includegraphics[width=.45\textwidth]{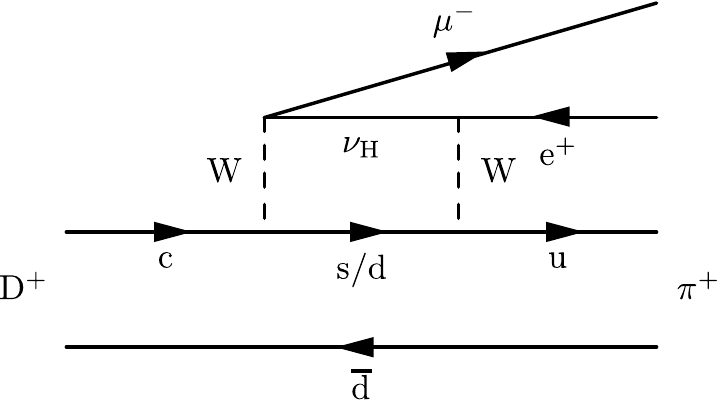}}
\caption{Examples for introduction of lepton flavour violating in meson decays}
\label{fig:hadlfvbsm}
\end{figure}

Two examples how to introduce sizeable lepton flavour violation are multi Higgs
extensions by means of new scalar particles (see e.\,g.\ \cite{d44_1461}) as in
the diagram in Fig.~\ref{fig:phis} or by means of heavy neutrinos as introduced
in low scale seesaw models (e.\,g.\ \cite{D73_074011}) which couple to
electrons and muons as shown in Fig.~\ref{fig:hadlfvbsmtwo}. Other ways to
embed LFV in the standard model are given e.\,g.\ in \cite{leptoquark}.

Particularly interesting about LFV in $\PB$ decays compared to $\PD$ decays is
that the $\PB$ mass is high enough to produce a $\APtauon, \Pmuon$ pair in the
final state. For new physics introduced in a Higgs coupling, this final state
is most sensitive due to the high masses, and thereby Higgs couplings of the
leptons involved.

\begin{figure}
\centering
\subfloat[][standard model with neutrino oscillation]{\includegraphics[width=.4\textwidth]{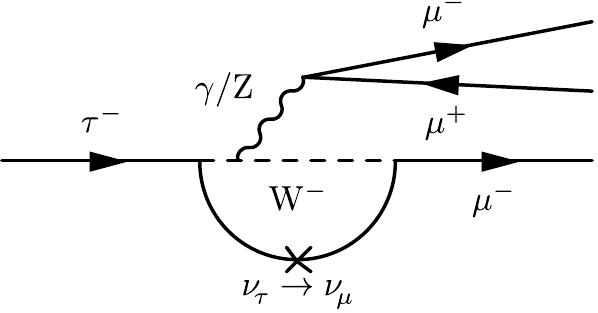}}\hspace{.05\textwidth}
\subfloat[][new physics with a charged Higgs]{\includegraphics[width=.37\textwidth]{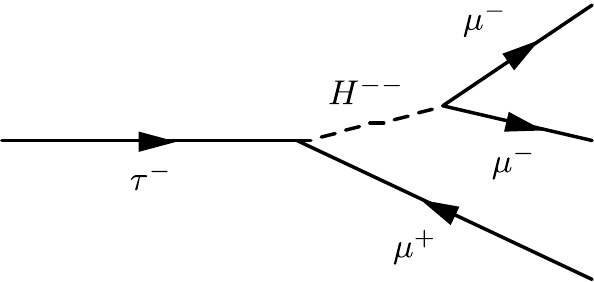}}
\caption{Feynman diagrams for $\Ptauon\to\Pmuon\Pmuon\APmuon$ in different models.}
\label{fig:tau23mutheory}
\end{figure}

Similarly the decay $\Ptauon\to\Pmuon\Pmuon\APmuon$ is not entirely forbidden,
but neutrino oscillation at loop level alone cannot bring the branching
fraction to an observable level. As presented in \cite{renga} the strongest
limits on LFV in lepton decays come from the $\Pmuon\to\Pelectron\Pphoton$
mode, the search for $\Ptauon\to\Pmuon\Pmuon\APmuon$ is particularly
interesting because some new physics models (e.\,g.\ Littlest Higgs
\cite{littlesthiggs}), as in Fig.~\ref{fig:tau23mutheory}, have enhanced lepton
flavour violating couplings to heavy leptons (favouring $\Ptau$ over $\Pmu$
decays) and do not involve photon couplings and therefore enhance the three
lepton final state over the $\Plepton\Pphoton$ final state
\cite{tau23mutheory}.  Moreover to identify the character of new physics, a
search in both $\Plepton\to\Plepton^\prime\Pphoton$ and
$\Plepton\to\Plepton^\prime\Plepton^\prime\Plepton^\prime$ must be performed.

\section{Experimental Results}

The study of rare decays naturally needs large event samples, which, for $\PB$
and $\PD$ mesons, is available at the $\PB$ factories and at the LHC.  The most
stringent constraints on LFV and LNV in modes involving electrons come from
BaBar and Belle, while muonic final states are now best constrained by recent
LHCb measurements.

\subsection{Limits on Lepton Number Violation}

Decays of $\PBplus$, $\PDplus$, and $\PKplus$ mesons were used to search for
Majorana neutrinos of different masses. The mass difference of the decaying
meson and the final state lepton is the upper limit on the mass of the on-shell
neutrino which can be produced. Since the neutrino mass is the invariant mass
of the final state meson-lepton pair, the sum of their rest masses is the lower
limit on the accessible mass range.

The strongest limits on the lepton coupling $|V_{\Pe 4}|^2$ and $|V_{\Pmu
4}|^2$ to a fourth neutrino are in the low neutrino mass region between
$\unit{140}{\MeVovercsq}$ and $\unit{353}{\MeVovercsq}$ coming from searches
for the decays $\PKplus\to\APelectron\APelectron\Ppiminus$ and
$\PKplus\to\APmuon\APmuon\Ppiminus$ respectively. Couplings down to
$|V_{\Plepton 4}|^2 \lesssim 10^{-8}$ are thereby ruled out in the most
sensitive range. 

The currently most stringent limits on LNV in charm decays and thereby higher
neutrino masses have been obtained by the BaBar collaboration
\cite{babar-charm} shown in Tab.~\ref{tab:babar-charm}.

The extension of the search range to higher masses is only possible in $\PB$
decays, the enormous production cross section in hadron collisions makes the
LHC the optimal place for searches for LNV in $\PB$ decays.  LHCb recently
provided new results on the on-shell modes $\PBplus\to\Ppiminus\APmuon\APmuon$
and $\PBplus\to\PDsminus\APmuon\APmuon$, as well as the virtual modes
$\PBplus\to\PDplus\APmuon\APmuon$ and $\PBplus\to\PD^{*-}\APmuon\APmuon$
\cite{lhcb-Blnv2}.

Limits on the branching fraction are hereby set as a function of the neutrino
mass for the on-shell modes.  For comparison, assuming a flat distribution of
the decay products in phase space, the observed branching fraction is shown in
Tab.~\ref{tab:BLNV} along with the modes which are sensitive to virtual
Majorana neutrinos and previous measurements.

The first search for $\PBplus\to\PDzero\Ppiminus\APmuon\APmuon$ has been
performed by LHCb \cite{lhcb-Blnv2} and showed no excess over the background.
Since this channel involves an on-shell Majorana neutrino, the limit is given
as a function of the neutrino mass as well.

\begin{table*}
\centering
\caption{Current limits on lepton number violating charm (a) and bottom (b) meson decays.}
\subfloat[][charm decays\label{tab:babar-charm}]{
\begin{tabular}{lclr}
channel & limit & & \\\hline
 $\mathcal{B}(\PDplus \to\Ppiminus \APelectron  \APelectron) $     &  $<1.9\times 10^{-6}$     &  @$90\,\%$ CL     &\cite{babar-charm} {\footnotesize{BaBar}} \\
 $\mathcal{B}(\PDplus \to\Ppiminus \APmuon      \APmuon)     $     &  $<2.0\times 10^{-6}$     &  @$90\,\%$ CL     &\cite{babar-charm} {\footnotesize{BaBar}} \\
 $\mathcal{B}(\PDplus \to\Ppiminus \APmuon      \APelectron) $     &  $<2.0\times 10^{-6}$     &  @$90\,\%$ CL     &\cite{babar-charm} {\footnotesize{BaBar}} \\
 $\mathcal{B}(\PDsplus\to\Ppiminus \APelectron  \APelectron) $     &  $<4.1\times 10^{-6}$     &  @$90\,\%$ CL     &\cite{babar-charm} {\footnotesize{BaBar}} \\
 $\mathcal{B}(\PDsplus\to\Ppiminus \APmuon      \APmuon)     $     &  $<14 \times 10^{-6}$     &  @$90\,\%$ CL     &\cite{babar-charm} {\footnotesize{BaBar}} \\
 $\mathcal{B}(\PDsplus\to\Ppiminus \APmuon      \APelectron) $     &  $<8.4\times 10^{-6}$     &  @$90\,\%$ CL     &\cite{babar-charm} {\footnotesize{BaBar}} \\
 $\mathcal{B}(\PDplus \to\PKminus  \APelectron  \APelectron) $     &  $<0.9\times 10^{-6}$     &  @$90\,\%$ CL     &\cite{babar-charm} {\footnotesize{BaBar}} \\
 $\mathcal{B}(\PDplus \to\PKminus  \APmuon      \APmuon)     $     &  $<10 \times 10^{-6}$     &  @$90\,\%$ CL     &\cite{babar-charm} {\footnotesize{BaBar}} \\
 $\mathcal{B}(\PDplus \to\PKminus  \APmuon      \APelectron) $     &  $<1.9\times 10^{-6}$     &  @$90\,\%$ CL     &\cite{babar-charm} {\footnotesize{BaBar}} \\
 $\mathcal{B}(\PDsplus\to\PKminus  \APelectron  \APelectron) $     &  $<5.2\times 10^{-6}$     &  @$90\,\%$ CL     &\cite{babar-charm} {\footnotesize{BaBar}} \\
 $\mathcal{B}(\PDsplus\to\PKminus  \APmuon      \APmuon)     $     &  $<13 \times 10^{-6}$     &  @$90\,\%$ CL     &\cite{babar-charm} {\footnotesize{BaBar}} \\
 $\mathcal{B}(\PDsplus\to\PKminus  \APmuon      \APelectron) $     &  $<6.1\times 10^{-6}$     &  @$90\,\%$ CL     &\cite{babar-charm} {\footnotesize{BaBar}} \\
 $\mathcal{B}(\PLambdac\to\APproton\APelectron  \APelectron) $     &  $<2.7\times 10^{-6}$     &  @$90\,\%$ CL     &\cite{babar-charm} {\footnotesize{BaBar}} \\
 $\mathcal{B}(\PLambdac\to\APproton\APmuon      \APmuon)     $     &  $<9.4\times 10^{-6}$     &  @$90\,\%$ CL     &\cite{babar-charm} {\footnotesize{BaBar}} \\
 $\mathcal{B}(\PLambdac\to\APproton\APmuon      \APelectron) $     &  $<16 \times 10^{-6}$     &  @$90\,\%$ CL     &\cite{babar-charm} {\footnotesize{BaBar}} \\
\end{tabular}
}
\hspace{.05\textwidth}
\subfloat[][bottom decays\label{tab:BLNV}]{
\begin{tabular}{lclrl}
channel & limit & & \\\hline
 $\mathcal{B}(\PBminus\to\Ppi^{+}\Pelectron\Pelectron) $    &  $<2.3\times 10^{-8}$    &  @$90\,\%$ CL      &\cite{babar-Blnv}  &{\footnotesize{BaBar}} \\
 $\mathcal{B}(\PBminus\to\PK^{+}\Pelectron\Pelectron)  $    &  $<3.0\times 10^{-8}$    &  @$90\,\%$ CL      &\cite{babar-Blnv}  &{\footnotesize{BaBar}} \\
 $\mathcal{B}(\PBminus\to\PK^{*+}\Pelectron\Pelectron) $    &  $<2.8\times 10^{-6}$    &  @$90\,\%$ CL    & \cite{cleo-Blnv}    &{\footnotesize{CLEO}}  \\
 $\mathcal{B}(\PBminus\to\Prho^{+}\Pelectron\Pelectron)$    &  $<2.6\times 10^{-6}$    &  @$90\,\%$ CL    & \cite{cleo-Blnv}    &{\footnotesize{CLEO}}  \\
 $\mathcal{B}(\PBminus\to\PD^{+}\Pelectron\Pelectron)  $    &  $<2.6\times 10^{-6}$    &  @$90\,\%$ CL      & \cite{belle-Blnv} &{\footnotesize{Belle}} \\
 $\mathcal{B}(\PBminus\to\PD^{+}\Pelectron\Pmuon)      $    &  $<1.8\times 10^{-6}$    &  @$90\,\%$ CL      & \cite{belle-Blnv} &{\footnotesize{Belle}} \\
 $\mathcal{B}(\PBminus\to\Ppi^{+}\Pmuon\Pmuon)         $    &  $<1.3\times 10^{-8}$    &  @$95\,\%$ CL    & \cite{lhcb-Blnv2}   &{\footnotesize{LHCb}}  \\
 $\mathcal{B}(\PBminus\to\PK^{+}\Pmuon\Pmuon)          $    &  $<5.4\times 10^{-7}$    &  @$95\,\%$ CL    &\cite{lhcb-Blnv}     &{\footnotesize{LHCb}}  \\
 $\mathcal{B}(\PBminus\to\PK^{*+}\Pmuon\Pmuon)         $    &  $<4.4\times 10^{-6}$    &  @$90\,\%$ CL    & \cite{cleo-Blnv}    &{\footnotesize{CLEO}}  \\
 $\mathcal{B}(\PBminus\to\Prho^{+}\Pmuon\Pmuon)        $    &  $<5.0\times 10^{-6}$    &  @$90\,\%$ CL    &\cite{cleo-Blnv}     &{\footnotesize{CLEO}}  \\
 $\mathcal{B}(\PBminus\to\PD^{+}\Pmuon\Pmuon)          $    &  $<6.9\times 10^{-7}$    &  @$95\,\%$ CL    & \cite{lhcb-Blnv2}   &{\footnotesize{LHCb}}  \\
 $\mathcal{B}(\PBminus\to\PD^{*+}\Pmuon\Pmuon)         $    &  $<2.4\times 10^{-6}$    &  @$95\,\%$ CL    &  \cite{lhcb-Blnv2}  &{\footnotesize{LHCb}}  \\
 $\mathcal{B}(\PBminus\to\PDs^{+}\Pmuon\Pmuon)         $    &  $<5.8\times 10^{-7}$    &  @$95\,\%$ CL    &  \cite{lhcb-Blnv2}  &{\footnotesize{LHCb}}  \\
 $\mathcal{B}(\PBminus\to\PDzero\Ppiplus\Pmuon\Pmuon)  $    &  $<1.5\times 10^{-6}$    &  @$95\,\%$ CL    &  \cite{lhcb-Blnv2}  &{\footnotesize{LHCb}} 
\end{tabular}
}
\end{table*}

\begin{figure*}
\centering
\includegraphics[width=.45\textwidth]{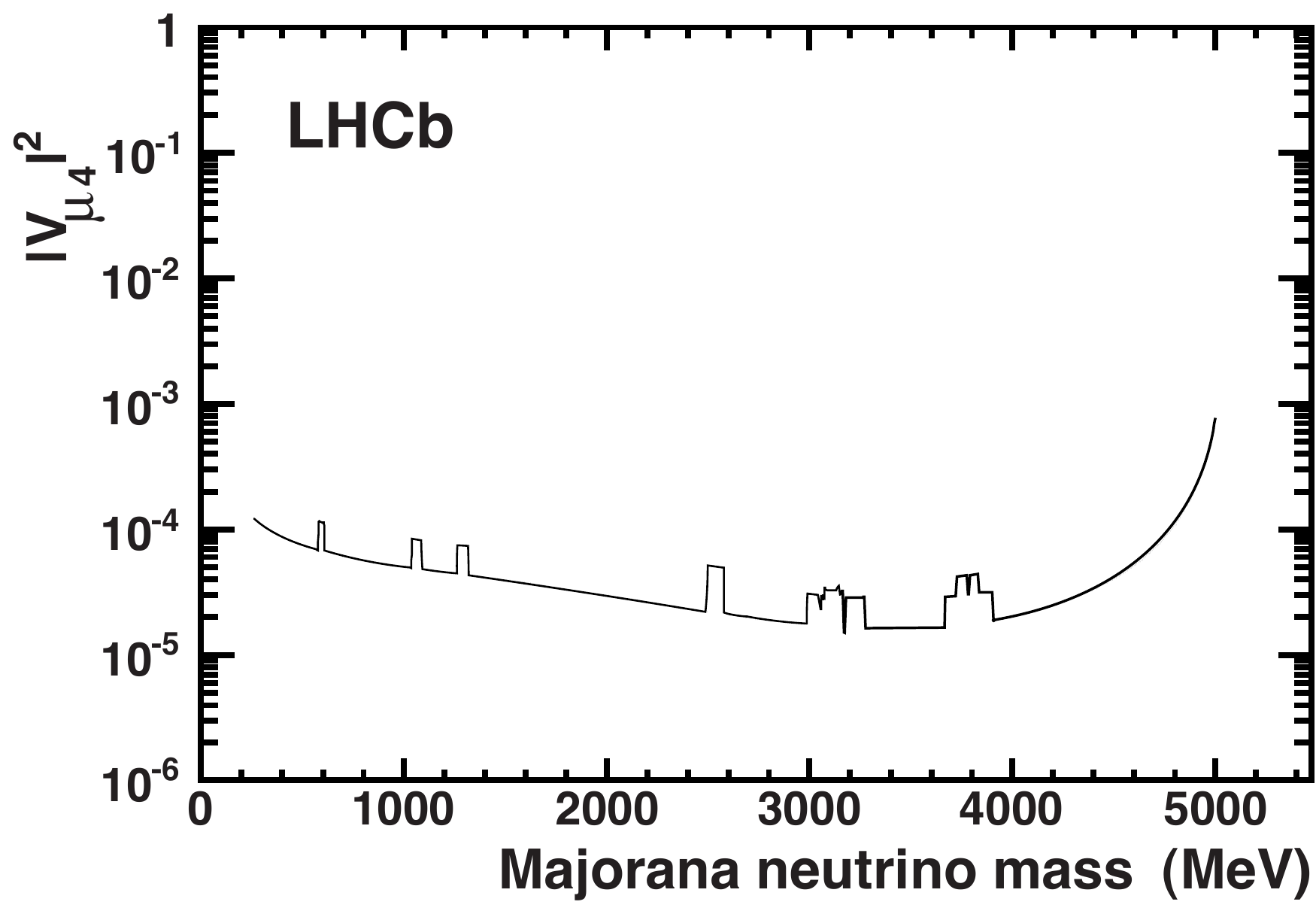}
\caption{Limit on $|V_{\mu 4}|^2$ from $\PBminus\to\Ppi^{+}\Pmuon\Pmuon$ measured by LHCb \cite{lhcb-Blnv2}.}
\label{fig:vmu4limit}
\end{figure*}

LHCb also provides the strongest limits on $|V_{\mu 4}|$ up to the $\PBplus$
mass considering these results come from $\PBplus\to\Ppiminus\APmuon\APmuon$,
shown in Fig.~\ref{fig:vmu4limit}.

\begin{figure*}
\centering
\includegraphics[width=.95\textwidth]{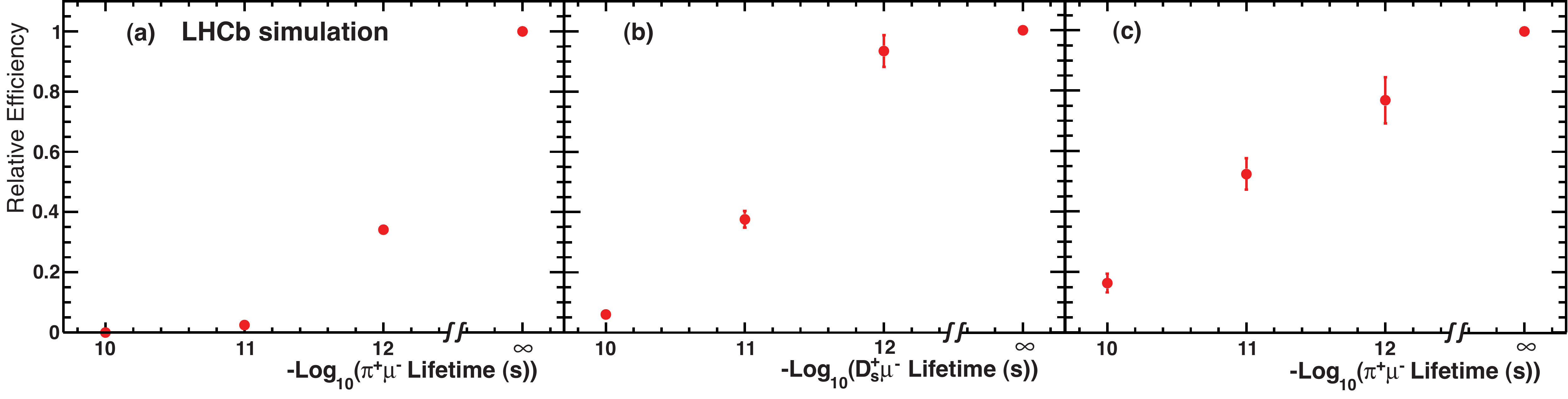}
\caption{Relative reconstruction efficiency as a function of the Majorana
neutrino lifetime. The branching fraction limits from~\cite{lhcb-Blnv2} have
been computed for the assumption of infinitively short lifetimes ($100\,\%$
relative efficiency). For longer lifetimes, the reconstruction efficiency
decreases and the observed limit has to be scaled down. The efficiencies are
given for (a) $\PBplus\to\Ppiminus\APmuon\APmuon$, (b)
$\PBplus\to\PDsminus\APmuon\APmuon$, and (c)
$\PBplus\to\PDzero\Ppiminus\APmuon\APmuon$.}
\label{fig:lifetime}
\end{figure*}

A natural way to search for a lepton number violating decay is to search for
same sign leptons from a common vertex. Thereby the analysis' implications on
an intermediate on-shell neutrino are only drawn correctly if the lifetime of
the neutrino is short enough not to degrade the reconstruction. To estimate how
the observed limits are to be understood in models with long lived heavy
neutrinos, LHCb also provides the relative reconstruction efficiency as a
function of the neutrino lifetime, shown in Fig.~\ref{fig:lifetime}.

\subsection{Limits on Lepton Flavour Violation}

The tightest constraints on lepton flavour violating processes in charm decays
have been found by the BaBar collaboration, listed in
Tab.~\ref{tab:babar-charmlfv}.  For bottom decays,
Tab.~\ref{tab:babar-bottomlfv} shows the recent results, involving $\Ptau$
leptons in the final state. Details are given in \cite{giovanni}. These results
improved the limit on the energy scale at which LFV can occur
\cite{LFV_energyscale} significantly.

The implication for new physics is that the energy scale for LFV effective
operators is pushed up from $\unit{2.2}{\TeV}$ to $\unit{11}{\TeV}$ or from
$\unit{2.6}{\TeV}$ to $\unit{15}{\TeV}$ for the $\Pbottom\to\Pdown$ and the
$\Pbottom\to\Pstrange$ transition respectively \cite{LFV_energyscale}.

\begin{table*}[b!ht]
\centering
\caption{Limits on lepton flavour violating hadron decays at $90\,\%$ confidence level. All listed limits from the BaBar collaboration.}
\subfloat[][charm decays\label{tab:babar-charmlfv}]{\begin{tabular}{lcr}
channel & limit  & \\\hline
 $\mathcal{B}(\PDplus \to\Ppiplus \APmuon      \Pelectron) $     &  $<3.6\times 10^{-6}$            &\cite{babar-charm}\\
 $\mathcal{B}(\PDplus \to\Ppiplus \APelectron      \Pmuon) $     &  $<2.9\times 10^{-6}$            &\cite{babar-charm}\\
 $\mathcal{B}(\PDsplus\to\Ppiplus \APmuon      \Pelectron) $     &  $< 20\times 10^{-6}$            &\cite{babar-charm}\\
 $\mathcal{B}(\PDsplus\to\Ppiplus \APelectron      \Pmuon) $     &  $< 12\times 10^{-6}$            &\cite{babar-charm}\\
 $\mathcal{B}(\PDplus \to\PKplus  \APmuon      \Pelectron) $     &  $<2.8\times 10^{-6}$            &\cite{babar-charm}\\
 $\mathcal{B}(\PDplus \to\PKplus  \APelectron      \Pmuon) $     &  $<1.2\times 10^{-6}$            &\cite{babar-charm}\\
 $\mathcal{B}(\PDsplus\to\PKplus  \APmuon      \Pelectron) $     &  $<9.7\times 10^{-6}$            &\cite{babar-charm}\\
 $\mathcal{B}(\PDsplus\to\PKplus  \APelectron      \Pmuon) $     &  $< 14\times 10^{-6}$            &\cite{babar-charm}\\
 $\mathcal{B}(\PLambdac\to\Pproton\APmuon      \Pelectron) $     &  $<19 \times 10^{-6}$            &\cite{babar-charm}\\
 $\mathcal{B}(\PLambdac\to\Pproton\APelectron      \Pmuon) $     &  $<9.9\times 10^{-6}$            &\cite{babar-charm}\\
\end{tabular}
}
\hspace{.05\textwidth}
\subfloat[][bottom decays\label{tab:babar-bottomlfv}]{
\begin{tabular}{lcr}
channel & limit  & \\\hline
$\mathcal{B}(\PBplus\to\PKplus\Ptauon\APmuon)$        & $<4.5\times 10 ^{-5}$ & \cite{babar-Blfv2}\\
$\mathcal{B}(\PBplus\to\PKplus\APtauon\Pmuon)$        & $<2.8\times 10 ^{-5}$ & \cite{babar-Blfv2}\\
$\mathcal{B}(\PBplus\to\PKplus\Ptauon\APelectron)$    & $<4.3\times 10 ^{-5}$ & \cite{babar-Blfv2}\\
$\mathcal{B}(\PBplus\to\PKplus\APtauon\Pelectron)$    & $<1.5\times 10 ^{-5}$ & \cite{babar-Blfv2}\\
$\mathcal{B}(\PBplus\to\Ppiplus\Ptauon\APmuon)$       & $<6.2\times 10 ^{-5}$ & \cite{babar-Blfv2}\\
$\mathcal{B}(\PBplus\to\Ppiplus\APtauon\Pmuon)$       & $<4.5\times 10 ^{-5}$ & \cite{babar-Blfv2}\\
$\mathcal{B}(\PBplus\to\Ppiplus\Ptauon\APelectron)$   & $<7.4\times 10 ^{-5}$ & \cite{babar-Blfv2}\\
$\mathcal{B}(\PBplus\to\Ppiplus\APtauon\Pelectron)$   & $<2.0\times 10 ^{-5}$ & \cite{babar-Blfv2}\\
$\mathcal{B}(\PBplus\to\Ppiplus\Pmu^{\pm}\Pe^{\mp})$  & $<1.7\times 10 ^{-7}$ & \cite{babar-Blfv}\\
\end{tabular}}
\end{table*}

The most recent result in this talk is the limit on LFV in
$\Ptauon\to\Pmuon\Pmuon\APmuon$ achieved by LHCb \cite{tau23muCONF}.  The
hadron collider environment introduces special experimental challenges compared
to the $\PB$ factories.

\paragraph{$\Ptau$ tag} At the $\PB$ factories, $\Ptau$ are produced in pairs.
A clean event selection therefore is to look at events with four tracks --
three from the signal candidate and one from a standard model one prong $\Ptau$
decay.  At the LHC the main source for $\Ptau$ is the leptonic
$\PDsminus\to\Ptauon\APnut$ decay \cite{tau23muCONF}.

\paragraph{Normalisation} The $\Ptau$ tag automatically provides the number of
produced $\Ptau$ which enter the analysis. Since the main production mode for
$\Ptau$ at the LHC does not provide any further charged particles, the number
of $\Ptau$ entering the analysis is not directly accessible. A normalisation to
allowed $\Ptau$ decays is not possible since they are indistinguishable from
more abundant $\PDplus$ decays with $\Ppizero$ in the decay chain.

\paragraph{Background} Having no production tag, background from events without
$\Ptau$, such as $\PB$ and $\PD$ cascade decays, is more severe in the LHCb
analysis than for the $\PB$ factories.

The main advantage of LHCb however is the huge production cross section for
$\Ptau$ from $\PDs$ decays. Considering the charm and bottom production cross
sections measured by LHCb \cite{LHCb-CONF-2010-013,Bprod} and the known
semileptonic branching fractions \cite{pdg}, about $8\times 10^{10}$ $\Ptau$
leptons have been produced at LHCb in 2011 compared to a total of $10^9$
$\Ptau$ pairs at the $\PB$ factories.

The analysis strategy of \cite{tau23muCONF} is similar to other rare decay
searches at LHCb. A loose cut based selection is applied to get a processable
data sample.  All events passing this selection are classified in a three
dimensional likelihood space. The discriminating variables are the invariant
mass of the $\Ptauon\to\Pmuon\Pmuon\APmuon$ candidate, a multivariate
classifier $\mathcal{M}_\text{3body}$ for the three body decay properties
(geometry, displacement, track quality, isolation, and kinematics), and a
multivariate classifier for the particle identification
$\mathcal{M}_\text{PID}$ (combining information from muon stations, RICH
detectors, and the calorimeter signature).  The latter classifiers use boosted
decision trees \cite{bdt} with adaptive boosting \cite{bdt_boosted} as
implemented by TMVA \cite{tmva}.

\begin{figure*}
\centering
\subfloat[][Distribution for simulated background and the simulated signal as a function of the 3 body decay classifier.]{\includegraphics[width=.44\textwidth]{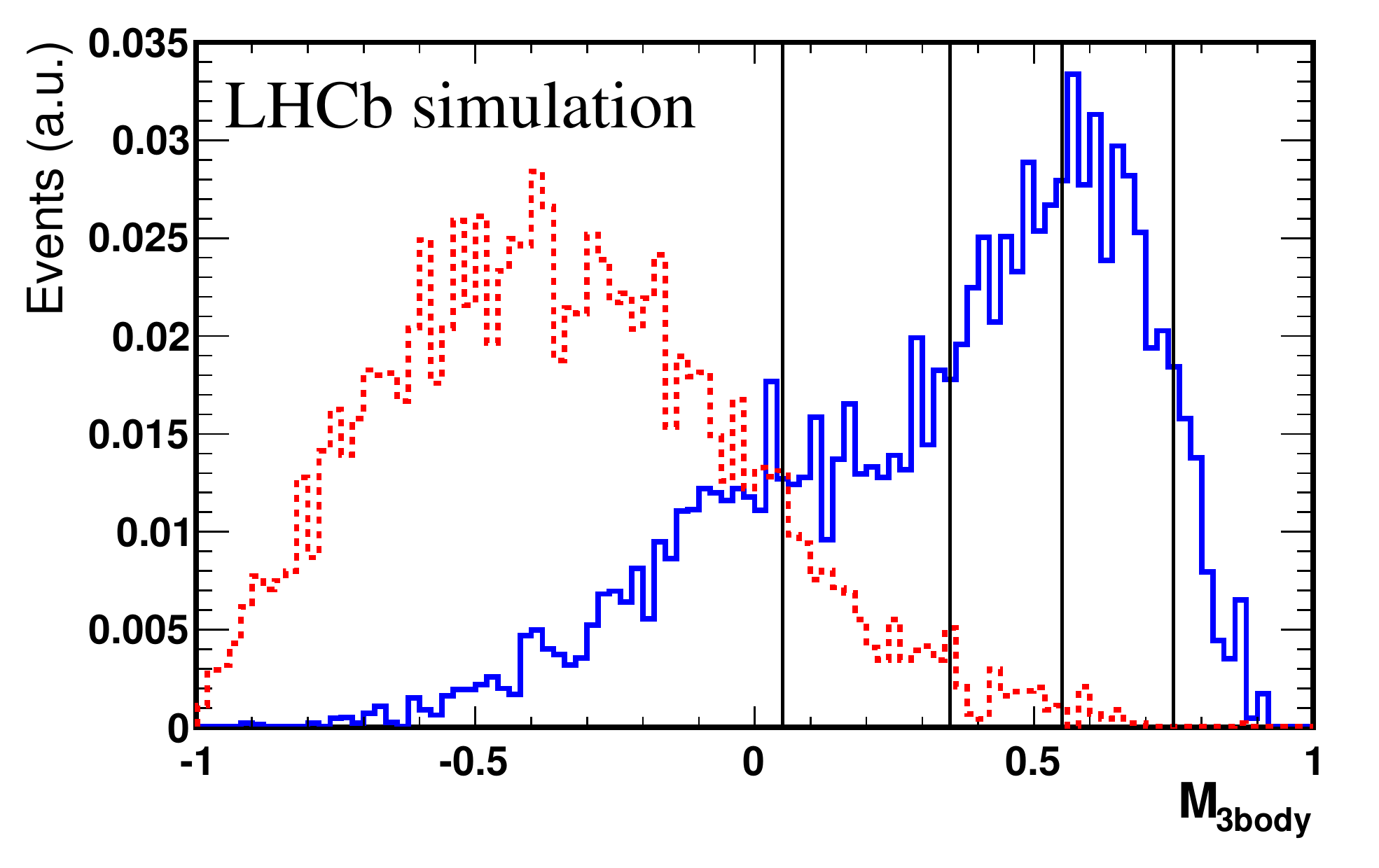}}
\hspace{.05\textwidth}
\subfloat[][Distribution for simulated background and the simulated signal as a function of the PID classifier.]{\includegraphics[width=.44\textwidth]{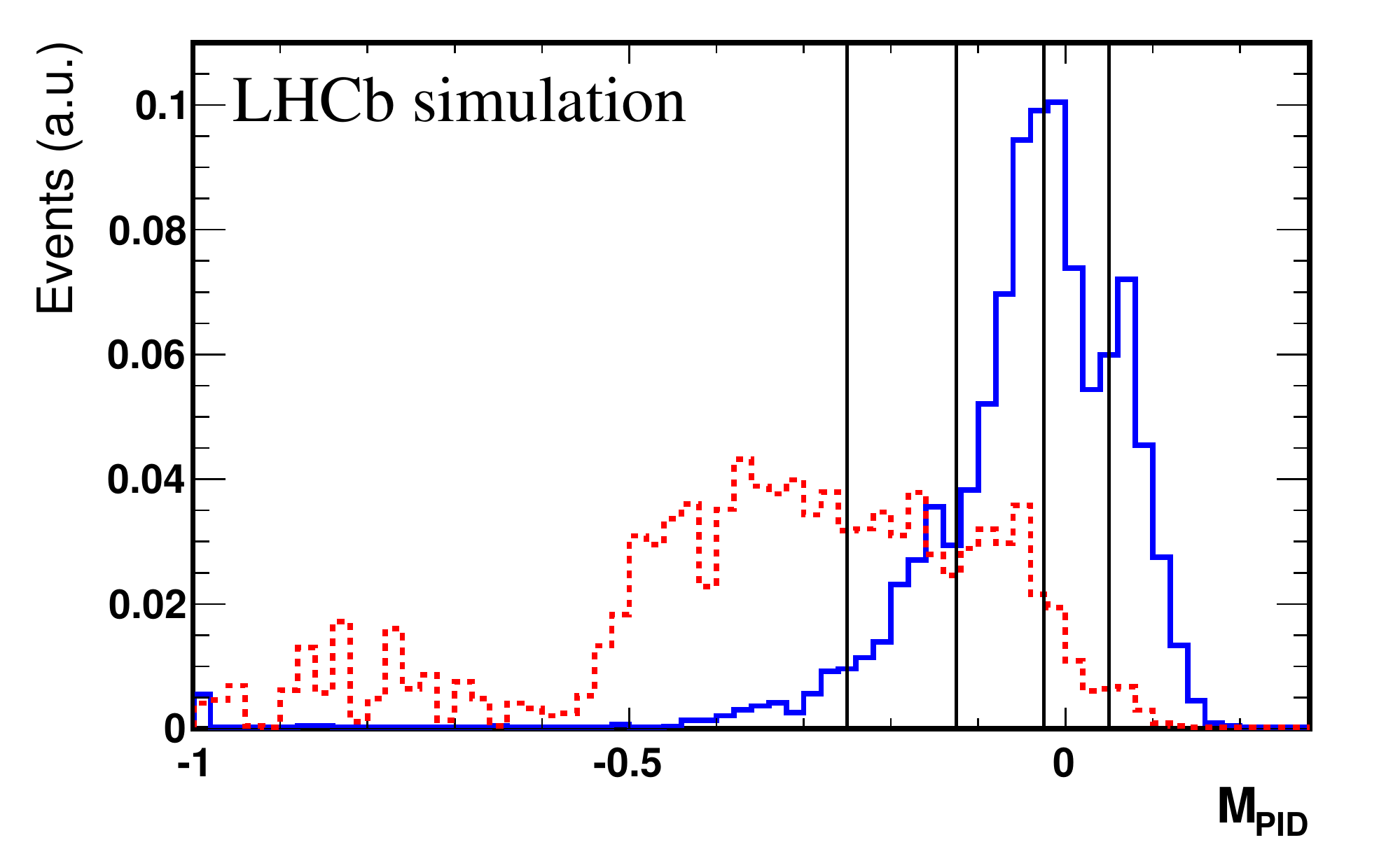}}
\caption{Distribution of signal events in the two multivariate likelihoods for signal (blue / solid) and background (red / dashed).}
\label{fig:binning}
\end{figure*}

\begin{figure*}
\centering
\subfloat[][simulated signal\label{fig:massbinning}]{\includegraphics[width=.45\textwidth]{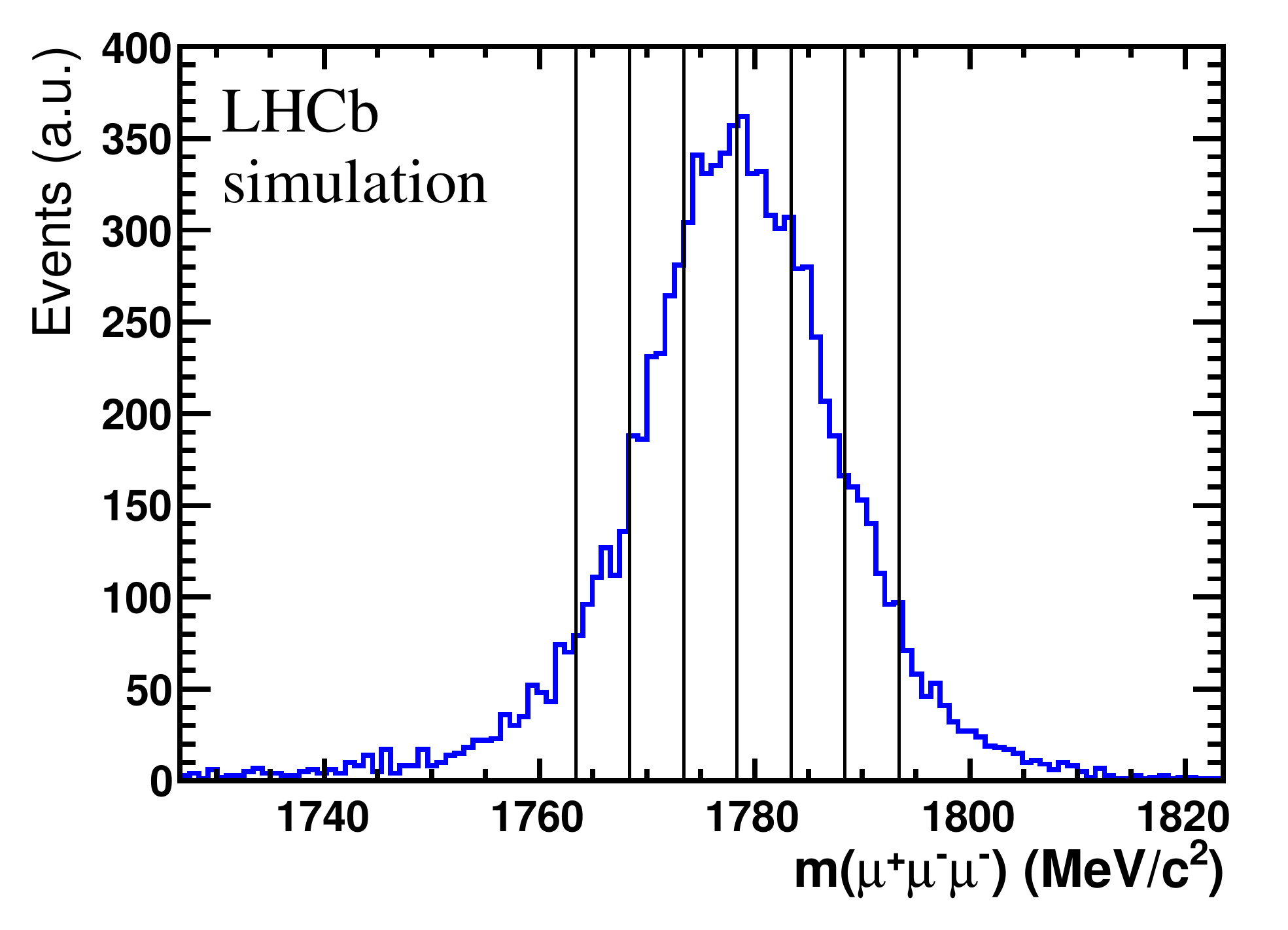}}
\hspace{.05\textwidth}
\subfloat[][observed events\label{fig:unblinded}]{\includegraphics[width=.45\textwidth]{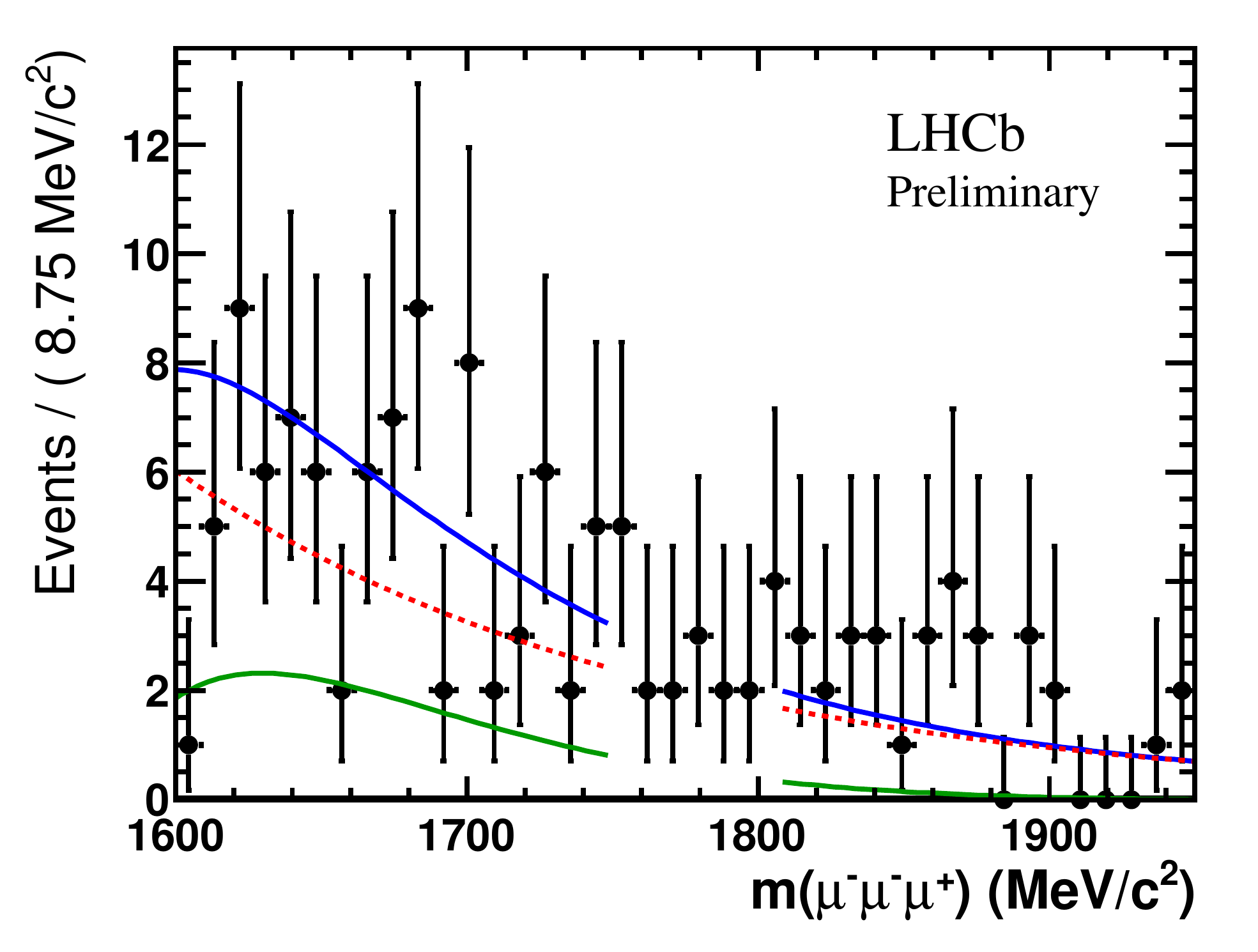}}
\caption{Invariant mass distribution for (a) simulated signal candidates and (b) observed events in the two highest $\mathcal{M}_\text{3body}$ and $\mathcal{M}_\text{PID}$ bins. The background fit ($\PDsplus\to\Peta\APmuon\Pnu$ in green / dotted; combinatorial in red / dashed; combined in blue / solid) is shown in the range which is used for the fit.}
\label{fig:tau_and_background_mass}
\end{figure*}

The signal efficiency of the multivariate classifiers as well as the invariant
mass resolution come from simulation and are calibrated on a control channel --
$\PDsplus\to\Ppiplus\Pphi(\APmuon\Pmuon)$ in the case for the three body
classifier and the invariant mass and $\PBplus\to\PJpsi(\APmuon\Pmuon)\PKplus$
for the particle identification.

The $\PDsplus\to\Pphi\Ppiplus$ calibration channel also serves as a
normalisation, since the branching fractions
$\mathcal{B}(\PDsplus\to\Pphi\Ppiplus)$ and
$\mathcal{B}(\PDsplus\to\APtauon\Pnut)$ are known -- yielding the number of
$\Ptau$ which have been produced in $\PDs$ decays. To determine the fraction of
$\Ptau$ from $\PDs$ decays, $f(\PDs)$, the bottom and charm cross sections
measured by LHCb \cite{LHCb-CONF-2010-013,Bprod}, as well as the branching
fractions of charm and bottom hadrons to $\Ptau$ are used. Hereby most of the
systematic uncertainties (e.\,g.\ luminosity measurement, reconstruction
efficiencies) cancel, i.\,e.\ $f(\PDs)$ is more accurately known than the
inclusive $\Ptau$ production cross section.  Contributions from gauge bosons or
Drell-Yan processes have been evaluated to be negligible.

Using the above normalisation as well as the efficiencies for selection,
reconstruction, and trigger the branching fraction can be written as follows:
\begin{align*}\mathcal{B}(\Ptauon\to\Pmuon\Pmuon\APmuon)&=
\frac{\mathcal{B}(\PDsplus\to\Pphi(\APmuon\Pmuon)\Ppiplus)}{\mathcal{B}(\PDsplus\to\APtauon\Pnut)}\times f(\PDs)\\
&\quad\times\frac{\varepsilon_\text{norm}}{\varepsilon_\text{sig}}\frac{N_{\Ptauon\to\Pmuon\Pmuon\APmuon}}{N_{\PDsplus\to\Pphi(\APmuon\Pmuon)\Ppiplus}}\end{align*}

where $\varepsilon_\text{norm}$ is the total efficiency to trigger, reconstruct
and select the normalisation decay and $\varepsilon_\text{sig}$ is the total
efficiency for the signal channel.

The dimuon decay of the $\Pphi$ is chosen to provide similar trigger and
particle identification properties compared to the signal being sought for. The
non resonant contribution from $\PDsplus\to\APmuon\Pmuon\Ppiplus$ decays was
found to be below $2\,\%$.

The three dimensional likelihood space is subdivided into 150 bins (five for
$\mathcal{M}_\text{3body}$ and $\mathcal{M}_\text{PID}$, and six for the
invariant mass) as shown in Fig.~\ref{fig:binning} and~\ref{fig:massbinning}.
The signal efficiency for each bin is evaluated from the calibration channels
and the background in each bin is estimated from the sidebands in the invariant
mass.

The background consists mainly of two components. Firstly combinatorial
background which is modelled by an exponential and secondly by
$\PDsplus\to\Peta(\APmuon\Pmuon\Pphoton)\APmuon\Pnum$ decays.
This physical background is not discriminated in the current analysis by either
$\mathcal{M}_\text{3body}$ or $\mathcal{M}_\text{PID}$ as it has the same
behaviour as the signal in all input quantities. Rejecting this decay will be
subject of future improvements.
%
%
%
It is modelled by an exponential multiplied by a second order polynomial for
which all shape parameters have been fixed on simulated events. The
normalisation is left free in the final fit within one standard deviation from
the expected yield which is determined using the normalisation channel and the
branching fractions $\mathcal{B}(\PDsplus\to\Peta\APmuon\Pnum),
\mathcal{B}(\Peta\to\APmuon\Pmuon\Pphoton)$, and
$\mathcal{B}(\PDsplus\to\Pphi\Ppiplus)$. For illustration the invariant mass
distribution and the combined fit for the combination of the two highest
$\mathcal{M}_\text{3body}$ and two highest $\mathcal{M}_\text{PID}$ bins is
shown in Fig.~\ref{fig:unblinded}.

For the final limit, all bins are combined using the CLs method
\cite{CLs1,CLs2}.
\begin{table}[tb]
\centering
\caption{Limits on the branching fraction for $\Ptauon\to\Pmuon\Pmuon\APmuon$ obtained by different experiments.}
\label{tab:tau23muresult}
\begin{tabular}{lclr}
collaboration & limit & \\\hline
Belle                                    & $<2.1\times 10^{-8}$ & @$90\,\%$ CL & \cite{tau23muBELLE} \\
BaBar                                    & $<3.3\times 10^{-8}$ & @$90\,\%$ CL & \cite{tau23muBABAR} \\
LHCb             & $<6.3 \times 10^{-8}$ & @$90\,\%$ CL & \cite{tau23muCONF} 
\end{tabular}
\end{table}
The observed limits at $90\,\%$ confidence level is $6.3\times 10^{-8}$, in
agreement with the expected limit for the absence of a signal ($8.2\times
10^{-8}$). Tab.~\ref{tab:tau23muresult} shows the comparison to the limits from
BaBar and Belle.

{~}

{~}

{~}

\section{Conclusion}

Hadron decays measured at the $\PB$ factories and at the LHC provide an
excellent and abundant probe to search for LNV and LFV. So far no signal has
been observed and only lower limits for the branching fractions are given.  The
$\PB$ factories have achieved high sensitivity and ruled out branching
fractions to the level of $10^{-5}$. The huge production cross section of $\PB$
mesons at the LHC furthermore enabled LHCb to improve the limits on LNV in
$\PB$ decays further.  Finally, the first LFV $\Ptau$ decay search performed at
a hadron collider has been performed.

{~}

{~}

{~}

{~}

{~}